# High-Level Design of Portable and Scalable FPGA Accelerators

*Extended Abstract*

Markus Weinhardt, Rainer Höckmann, Thomas Kinder
Osnabrück University of Applied Sciences, Osnabrück, Germany
*m.weinhardt | r.hoeckmann | t.kinder@hs-osnabrueck.de*

## I. INTRODUCTION

This paper presents our approach for making FPGA accelerators accessible to software (SW) programmers. It is intended as a starting point for collaborations with other groups pursuing similar objectives. We report on our current SAccO platform (**S**calable **Acc**elerator platform **O**snabrück) [1], [2] and the planned project extending this platform.

SAccO is used for accelerating parts of multi-process streaming applications implemented on a standard PC. In SAccO, the portable and scalable accelerators are implemented on standard FPGA boards connected via PCI-Express (PCIe) [3] and designed as parameterizable RT-level VHDL hardware (HW) components. To simplify application development, a fast and powerful PCIe controller capable of DMA was developed for Xilinx FPGAs, together with an efficient driver. The controller is combined with a high-level communication API which uses the same function calls for SW-SW communication via sockets and SW-HW communication via PCIe.[1] This API enables simple process integration and also allows for pure SW applications (on PCs without FPGA extension cards) without changing the processes' source code.

SAccO also comprises a method to automatically select a process' optimal degree of parallelism on an FPGA for a given hardware platform, i. e. to generate a hardware design which uses the available FPGA resources and communication bandwidth between the PC and the FPGA optimally. Hence SW programmers can easily adjust the VHDL components to new platforms of varying size, performance and cost.

Due to recent advances in high-level synthesis (HLS) [4] it now becomes feasible to replace the VHDL components by HLS-generated HW designs. This will be explored in the course of this new project. Then the SAccO PCIe interface and API can be used by SW engineers without HW design knowledge. However, even with HLS, a lot of difficult manual fine tuning is required. Different loop unrolling or tilling parameters result in different size/speed trade-offs, cf. [5]. Furthermore, e. g. the ROCCC2.0 HLS compiler [6] makes the user select the communication bandwidth for a HW kernel, a platform-specific parameter which is normally not exposed to SW programmers. Therefore we will combine HLS with an automatic HW/SW partitioning and optimization algorithm which selects these parameters.

The remainder of this paper is organized as follows: The next section describes the background of our approach and related work. Sect. III outlines the combination of SAccO with an HLS system and a new optimization method. Finally, Sect. IV summarizes the paper.

## II. BACKGROUND

The performance of today's standard PCs (based on the x86 architecture) is not sufficient for computation-intensive applications. However, using ASICs or expensive supercomputers is often infeasible. Off-the-shelf FPGA boards are a reasonably priced alternative for upgrading standard PCs. Using them, computation-intensive application kernels can be mapped to fast coprocessors and configured into the FPGAs, thereby almost reaching the performance levels of expensive specialized hardware.

Unfortunately, these FPGA boards have not yet been widely used for accelerating standard PC programs. One reason is the fact that designing digital circuits is much more difficult and time-consuming than developing software [7]. To overcome this productivity gap, high-level design tools must be used. HLS has been researched for a long time [8], [9] and recently received increased interest due to improved tools [4].

Furthermore, most hardware accelerators are hand-optimized ad-hoc solutions for one specific FPGA board. Using other FPGA boards requires time-consuming porting or complete rewriting. Hence, the high development effort is not worthwhile if, e. g. for cost reasons, FPGAs of different sizes and vendors shall be used in different workplaces of a company. The lack of portability and scalability has not been researched thoroughly though it is recognized as a major obstacle for widespread use of FPGAs [10]. There are only a few publications on portable and scalable FPGA design, e. g. [11]. They also suggest implementations with varying degrees of parallelism, but do not present a method for automatically determining the hardware parameters as in our approach.

Finally, the PCIe interfaces provided by the FPGA board vendors only provide base functionality on a low level [12], [13]. The *Speedy PCIe Core* [14] and the *RIFFA2.0* [15] projects provide high-level PCIe interfaces. However, they are only available under restricted licenses and do not use the same API for SW-SW and SW-HW communication.

---

[1]The SAccO SW and HW components are freely available upon request.

For these reasons, in the *HPVis* project at Osnabrück University of Applied Sciences[2], we devised methods for designing portable and scalable hardware accelerators and implemented the PCIe interface, resulting in the SAccO platform.

## III. AUTOMATIC HIGH-LEVEL PARTITIONING AND SCALING USING HLS

In the new approach, the applications are based on multi-process streaming SW as in the HPVis project. However, instead of manually porting selected processes to VHDL, we will use HLS for implementing some processes in FPGA hardware. This will require some specific extensions. E. g., the SAccO API for process communication over streaming channels must be integrated in the HLS system. For this purpose, several commercial and research HLS systems will be evaluated.

In SAccO [2], an optimal replication factor $R$ is automatically computed for *one* HW kernel (corresponding to one process). $R$ determines the optimal parallelization degree from the available FPGA resources and the given PCIe communication bandwidth which differ vastly for different FPGA devices and PC platforms.

This method will be generalized to optimize *all* processes of an application. Therefore, the processes $P$ and their communication channels $C$ are represented as a data-flow graph $DFG = (P, C)$. A combinatorial optimization algorithm will be devised which

- selects the nodes (processes) to be implemented on the FPGA (*hardware/software partitioning*) and
- selects a replication (unrolling/tiling) factor $R$ which determines the degree of parallelism used for each hardware node (*optimal scaling*).

As opposed to early HW/SW partitioning methods, e. g. [16], which maximized performance and minimized the number of data word transfers, this algorithm's objective is to maximize the throughput from the DFG's source to its sink node. Several interdependent constraints restrict the solution space:

- All software processes share the CPU performance.
- All hardware processes share the FPGA resources. For replication factors $R > 1$, the performance is increased, but so are the resource requirements.
- The process performance limits the throughput of its connected communication channels.
- The throughput of a channel is determined by the minimum of the channel bandwidth, the source's production rate and the sink's consumption rate.
- The channel throughput in turn limits the performance of its connected processes.
- All channels between SW and HW processes must share the limited PCIe bandwith.

[2]HPVis: High-Performance Processing and Visualization of High-Volume Data, cf. http://www.ecs.hs-osnabrueck.de/hpvis.html, supported by the European Regional Development Fund and the Lower Saxony State Government/Germany.

This problem is related to the maximum-flow problem but is more complicated since the flow depends on the processes' performance, and the bandwidth is variable (depending on $R$). An efficient solution algorithm (e. g. a maximum-flow or integer linear-programming algorithm) will be developed.

Apart from this algorithmic work, methods for extracting the system parameters (process performances, channel bandwidths etc.) by profiling or directly from the software compiler's and the HLS system's output must be developed. The scheduling of the channel communication over the SW/HW boundary will be implemented by a method similar to [17].

## IV. SUMMARY

This paper presented a summary of our work on the free SAccO platform which simplifies the integration and reuse of hardware kernels for software programmers. Next, we outlined how this platform can be combined with High-Level Synthesis systems and how an extended optimization algorithm performs hardware/software partitioning and hardware optimization.